\documentstyle[12pt]{article}
\font\it=cmti10 scaled\magstep 1

\newcommand{\newc}{\newcommand}
\newc{\ra}{\rightarrow}
\newc{\lra}{\leftrightarrow}
\newc{\beq}{\begin{equation}}
\newc{\eeq}{\end{equation}}
\newc{\barr}{\begin{eqnarray}}
\newc{\earr}{\end{eqnarray}}
\newc{\bed}{\begin{displaymath}}
\newc{\edd}{\end{displaymath}}
\begin{document}
\begin{flushright}
{IOA.304/94}\\
\end{flushright}
\vspace*{3cm}
\begin{center}
{\bf A NEW TREATMENT OF NEUTRINO}\\
{\bf  OSCILLATIONS IN MEDIUM}\\
\vspace*{2cm}
{\bf G.K. Leontaris and J. D. Vergados}\\
{\it Theoretical Physics Division} \\
{\it Ioannina University} \\
{\it GR-45110 Greece} \\
\vspace*{0.5cm}
\vspace*{0.5cm}
{\bf ABSTRACT} \\
\end{center}
\noindent

A new more rigorous and accurate method for treating neutrino oscillations
in the context of the MSW effect in a medium is proposed.  This leads to a
new type of resonance condition which for small mixing angles puts rather
stringent conditions on $E_{\nu}/\delta {m^2}$.  The implications on the
solar neutrino problem are discussed.

\vspace*{2cm}
\noindent
\begin{flushleft}
IOA-304/94\\

April 1994
\end{flushleft}
\thispagestyle{empty}
\vfill\eject
\setcounter{page}{1}


\noindent

One of the most important questions of the physics beyond the standard
model is the problem of neutrino masses.  Furthermore if the neutrinos are
massive, the neutral leptons produced in weak interactions are  not
stationary.  They are linear combinations of the neutrino mass eigenstates
(neutrino mixing).  Even though at present there is no theory which can
predict the mass and mixing of the neutrinos, most Grand Unified Models
predict small masses and mixing.  If the neutrinos are almost  degenerate,
neutrino oscillation experiments $^{\cite{1,2}}$ are the best candidates
to measure small $\delta {m^2}$ (from $1eV^2$ down to $10^{-10} eV^2)$.
Furthermore, neutrino oscillations may explain the solar neutrino problem
$^{\cite{3,4,5,6}}$,  i.e.  the apparent
reduction of the $\nu_e$ flux at earth compared to that predicted by the
standard solar model $^{\cite{7,8}}$ (SSM).  The mechanism of neutrino
oscillations however, is not effective if the neutrino mixing is small.
It has been observed, though, that under the conditions of high density
encountered in the sun's interior the oscillation can be enhanced due to
the MSW effect $^{\cite{9,10}}$.  In other words small mixing angles can be
converted into large effective mixing angles due to the resonant scattering
of $\nu_e$ neutrinos by electrons. Calculations of neutrino oscillations
involving the MSW effect have hitherto  involved the  following steps:

1) Convert the evolution equation from flavor space into neutrino mass
eigenstate basis by locally diagonalizing
the space dependent Hamiltonian$^{\cite{1,11}}$.

2) Ignore the  transitions between the mass eigenstates
(adiabatic approximation) or treat such transitions in perturbation
treatment.  In the last case one assumes that $^{\cite{12}}$.
\beq
\frac{\delta E_{\nu}/2 }{d\vartheta_m/dx}  \ll 1
\label{eq:1}
\eeq
where $\delta E_{\nu}$ is the difference in neutrino energies and
 $\frac{d\vartheta_m}{dx}$ the variation of the mixing angle
 in the medium with distance.  In
particular the above equation must be true at the  resonance point.

In the present paper  we will provide an exact solution which does not go
through the local neutrino mass eigenstates.  Our method is quite simple
and offers itself to a simple interpretation.

For illustration purposes we will exhibit our method in the case of two
generations but it can easily be extended to any number of generations.

Following standard procedure $^{\cite{11}}$
we can write the neutrino state at any time as
\bed
|\nu (t)> = a_e(t) |\nu_e> + a_{\alpha}(t) |\nu_{\alpha}>
\edd
where $|\nu_e>$  is the  electron neutrino and $\nu_{\alpha}$ any other
flavor (e.g. $\nu_{\mu}$).  The amplitudes $a_e(t)$ and
$a_{\alpha}(t)$  satisfy the evolution equation $^{\cite{9}}$
\begin{eqnarray}
\imath \frac{d}{dt}
\left(\begin{array}{c} a_e \\ a_{\alpha} \end{array}
 \right) = {\cal H}(t) \left(\begin{array}{c} a_e \\ a_{\alpha}
\end{array}
 \right). \label{eq:2}
\end{eqnarray}
Since $x=ct$, the above equation can be written in term of $x$.  In the
presence of matter it can be shown that ${\cal H}$ can be cast in the
form $^{\cite{1,13}}$
\begin{eqnarray}
{\cal H}
=\left[\begin{array}{cc}E_{\nu} - \frac{\pi}{\ell} cos 2\vartheta +
\frac{2\pi}{\ell_0(x)} &\frac{\pi}{\ell} sin2\vartheta \\
\frac{\pi}{\ell} sin2\vartheta &E_{\nu}+ \frac{\pi}{\ell} cos2\vartheta
\end{array}
 \right] \label{eq:3}
\end{eqnarray}
where
\beq
\ell = \frac{4\pi E_{\nu}}{\delta {m^2}} = 2.476 km \frac{(E_{\nu}/1
MeV)}{(\delta m^2/1 eV^2)} , \,\,\,\,\delta m^2 = m^2_{\alpha} - m^2_e
\label{eq:4}
\eeq
$\vartheta$ is the usual (vacuum) mixing angle and $\ell_0(x)$ takes into
account the fact that the charged current interaction between $\nu_e$ and
electrons causes a shift in the electron neutrino energy. $\ell_0(x)$
takes the form
\beq
\ell_0(x) = \frac{4\pi}{2\sqrt{2} G_F \rho_e(x)}
\label{eq:5}
\eeq
where $G_F= 1.16636\times 10^{-5}GeV^{-2}$ is the Fermi coupling constant,
and $\rho_e(x)$ in the number of electrons per unit volume at distance $x$
from the sun's center which is assumed to be spherically symmetric.
Equation  (\ref{eq:2}) can easily be integrated to yield
\begin{eqnarray}
\left(\begin{array}{c} a_e \\ a_{\alpha} \end{array}
 \right) = exp (-\imath {\cal A}) \left(\begin{array}{c} a_e \\
a_{\alpha} \end{array}
 \right)_0 \label{eq:6}
\end{eqnarray}
where $\left(\begin{array}{c} a_e \\ a_{\alpha}
\end{array}
 \right)_0  $is the initial solution of the differential equation and
the matrix ${\cal A}$ is given by
\begin{eqnarray}
{\cal A }=  \frac{\pi(x-x_0)}{\ell}
\left[\begin{array}{cc} 2\xi -cos 2\vartheta  &sin2\vartheta \\
sin2\vartheta &cos2\vartheta
\end{array}
 \right] \label{eq:7}
\end{eqnarray}
where
\beq
\xi = \frac{\ell}{(x-x_0)} \int^x_{x_0}  \frac{dx^\prime}{\ell_0(x^\prime)}
\label{eq:8}
\eeq
The diagonal matrix with elements $E_{\nu}(x-x_0)$ has been omitted as
irrelevant.

Using the well known fact that the minimum polynomial associated with the
matrix ${\cal A}$ is of degree one, which permits $exp (-\imath {\cal A})$
to be written as a
linear combination of the identity matrix and ${\cal A}$ $^{\cite{14}}$,
 plus the fact that
\bed
|\nu(x_0)> = |\nu_e>
\edd
it is straightforward to show that
\beq
{\cal P}(\nu_e \ra \nu_e) = 1 - \frac{\ell^2_m}{\ell^2}
sin^2 2\vartheta sin^2
\pi \frac{(x-x_0)}{\ell_m} \label{eq:9}
\eeq
where
\begin{eqnarray}
\ell_m &=&\ell_m(x,x_0)\\ \nonumber
& =& \frac{\ell} {\sqrt{1+\xi^2-2\xi cos2\vartheta} }\\ \nonumber
&=&
\frac{\ell} {{\sqrt{(\xi -cos2\vartheta)^2 + sin^22\vartheta}}}
 \label{eq:10}
\end{eqnarray}
$\ell_m$ can be interpreted as the oscillation length.  In the presence
of matter, however, $\ell_m$ is a function not only of  the combination
($E_{\nu}/\delta m^2$),
but also of $\vartheta ,x$ (detector point) and $x_0$ (source point).
One can also talk about an effective mixing angle $\vartheta_m$ defined by
\beq
sin 2\vartheta_m (x,x_0) =
\frac{\ell_m(x,x_0)}{\ell} sin2\vartheta
 \label{eq:11}
\eeq
{}From  eq.(\ref{eq:10}) it can easily be seen that the maximum oscillation
probability occurs when
\beq
\xi = cos 2\vartheta \,\, ,\,\,\,\,\,\, (``resonance"\,\,\, condition)
 \label{eq:12}
\eeq
In this  case
\beq
sin 2\vartheta_m (x,x_0) = 1,\,\,\,\, \ell_m(x,x_0) =\ell/ sin2\vartheta
 \label{eq:13b}
\eeq
while eq.(\ref{eq:9}) becomes
\beq
{\cal P}(\nu_e \ra \nu_e) = 1 - sin^2
 \{\pi\frac{x-x_0}{\ell/sin2\vartheta}\}
 \label{eq:13a}
\eeq
The above condition (\ref{eq:12}) is reminiscent but not similar to
the usual resonance condition (see eqs (\ref{eq:21a}) and
(\ref{eq:21b}) below)  which occurs at some appropriate point $x_R$.  In
our exact treatment the ``resonance" condition (eq. (\ref{eq:12}))
involves both the initial $(x_0)$ and the final $(x)$ positions.  It is
affected by the cumulative effect of the density of the  medium and  not
by its value at some appropriate point $x_R$.  It is a sort of
``{\underline {global resonance" condition}}.
For numerical calculations we will cast $\xi$ in the form
\beq
\xi = \frac{t_0}{(x-x_0)} \int^x_{x_0}
 \frac{\rho_e(x^\prime)}{\rho_0}
dx^\prime
 \label{eq:14}
\eeq
where
\beq
t_0 = \frac{2\sqrt{2} G_F E_{\nu} \rho_0}{\delta m^2}
 \label{eq:15}
\eeq
with $\rho_0$ the sun's density at some suitable point (e.g. at its
center).  The gross features  of the resonance are affected by $t_0$, while
its details depend on the specific from of $\rho_e(x)/\rho_0$.  The value
of $t_0$ which is consistent with the present solar neutrino data, lies
in the range\cite{15}
\beq
0.5 \leq t_0 \leq 50
 \label{eq:16}
\eeq
This range maybe enlarged however  ($\sim 0.2-60$), if also  uncertainties
on the density $\rho_0$ are also included$^{\cite{12}}$. The resonance
condition imposes constraints  on the parameters $x_0, \delta m^2$ and
$E_{\nu}$($x$ is assumed to be fixed, i.e. $x=$  {\it sun - earth }
distance).  The half maximum width is given  by
\beq
\Gamma = 2sin2\vartheta
 \label{eq:17}
\eeq
which for small mixing angle puts stringent constraints on $t_0$ or
equivalently on the allowed values of $E_{\nu}/\delta m^2$.

Before proceeding  further with the discussion of our results we will
compare  the above new formulas with  those  which have been obtained
with the traditional approach.  By diagonalizing the matrix of eq.
(\ref{eq:3}) for each value of $x$  we obtain the
eigenvalues$^{\cite{1,13} }$
\beq
\lambda_{\pm} = E_{\nu} + \frac{\pi}{\ell_0} \pm \frac{\pi}{\ell_m(x)}
\label{eq:18}
\eeq
and the eigenvectors
\begin{eqnarray}
|\nu_L(x)\succ = cos\vartheta_m(x) |\nu_e\succ -
sin\vartheta_m(x) |\nu_\alpha\succ \label{eq:20a}\\
 |\nu_H(x)\succ = sin\vartheta_m(x) |\nu_e\succ +
cos\vartheta_m(x) |\nu_\alpha\succ  \label{eq:20b}
\end{eqnarray}
with
\begin{eqnarray}
\ell_m(x) = \ell/[1-2cos 2\vartheta \frac{\ell}{\ell_0(x)} +
(\frac{\ell}{\ell_0(x)})^2]^{1/2}\\
 sin 2 \vartheta_m(x) = sin 2\vartheta/[1-2cos 2\vartheta
\frac{\ell}{\ell_0(x)} + (\frac{\ell}{\ell_0(x)})^2]^{1/2}
\end{eqnarray}
The resonance in this approach occurs when the diagonal elements of
${\cal H}$ of eq.(\ref{eq:3}) are equal $^{\cite{9,10}}$,
i.e. at a point $x_R$  in the medium such that
\beq
\ell_0(x_R) cos 2\vartheta = \ell . \label{eq:21}
\eeq
At the resonance one finds
\begin{eqnarray}
sin 2\vartheta_m(x_R) = 1 \label{eq:21a}\\
\ell_m (x_R) = \ell/ sin2\vartheta  \label{eq:21b}
\end{eqnarray}
 Then by writing
\beq
|\nu(x)> = a_L(x) |\nu_L(x)> + a_H |\nu_H(x)> \label{eq:23}
\eeq
we obtain the following evolution equation
\begin{eqnarray}
\imath \frac{d}{dt}
\left(\begin{array}{c} a_L(x) \\ a_H(x) \end{array}
 \right) ={\cal B}(x) \left(\begin{array}{c} a_L(x) \\ a_H(x)
\end{array}
 \right) \label{eq:24}
\end{eqnarray}
where
\begin{eqnarray}
{\cal B} = \left[\begin{array}{cc} \lambda_-
&-\imath \frac{d\vartheta_m}{dx} \\
\imath \frac{d\vartheta_m}{dx} &\lambda_+
\end{array}
 \right] \label{eq:25}
\end{eqnarray}
Following a procedure analogous to that of our new method outlined above,
the oscillation probability takes the form
\begin{eqnarray}
{\cal P}(\nu_e \ra \nu_e) &=& \frac{1}{2}
\{1 + cos2\vartheta_m(x)
cos2\vartheta_m(x_0) \frac{\beta^2 + \gamma^2
cos(2\sqrt{\beta^2+\gamma^2})}{\beta^2 + \gamma^2}  \nonumber \\
&+& sin2\vartheta_m(x) sin2\vartheta_m(x_0) cos(2 \sqrt{\beta^2+\gamma^2})
\nonumber \\
&+& sin(2\vartheta_m(x) - 2\vartheta_m(x_0)) \frac{\gamma}
{\sqrt{\beta^2+\gamma^2}}
 sin(2\sqrt{\beta^2+\gamma^2}) \} \label{eq:26}
\end{eqnarray}
where
\beq
\gamma=\vartheta_m(x) - \vartheta_m(x_0)
\eeq
and
\beq
\beta = \pi \int^{x}_{x_0} \frac{dx}{\ell_m(x)} \label{eq:27}
\eeq
If the sun's density is discontinuous at its surface, we get
\beq
\beta = \pi \left[\int^{x_s}_{x_0} \frac{dx}{\ell_m(x)} +
\frac{x-x_s}{\ell}\right]\label{eq:28}
\eeq
where $x_s\equiv R_{\odot}$ is the sun's radius.
We notice that the quantities
$\vartheta_m(x)$ (and $\gamma $) and $\ell_m(x)$ are not independent
quantities but they are given by eqs(\ref{eq:21a} and \ref{eq:21b}).
In the special case $\gamma \ll \beta$ (adiabatic approximation) we
 obtain
\beq
{\cal P}(\nu_e\rightarrow \nu_e) =
 \frac{1}{2}
\{1+cos(2\vartheta_m(x) - 2\vartheta_m(x_0))\}
- sin2\vartheta_m(x) sin2\vartheta_m(x) sin^2\beta \label{eq:29}
\eeq
which coincides with the old expression$^{\cite{1,12}}$.
Furthermore in the absence of matter effects i.e. when $\vartheta_m(x) =
\vartheta_m(x_0) = \vartheta$ we obtain once again
the well known formula $^{\cite{1,2}}$
\beq
{\cal P}(\nu_e \rightarrow \nu_e) = 1 - sin^2 2\vartheta
sin^2 \{\pi\frac{x-x_0}{\ell}\}\label{eq:30}
\eeq
On the other hand if $\vartheta_m(x) = \vartheta$ and
$sin2\vartheta_m (x_0 ) = sin2\vartheta (x_R) = 1$
(i.e. the resonance condition occurs at $x_0$) one
obtains for small values $^{\cite{13}}$ of $\vartheta $
\beq
{\cal P}(\nu_e\rightarrow \nu_e) \cong \frac{1}{2} -
sin2\vartheta sin^2 \beta  \simeq \frac {1}{2}
\label{eq:31}
\eeq
The comparison of our treatment with the non-adiabatic treatment of the
earlier perturbative calcultations $^{\cite{12,16}}$ is not obvious.
 Level crossings $^{\cite{17}}$ etc do not enter in our non-perturbative
treatment which avoids the intermediate step of the eigenstates of
eqs.(\ref{eq:20a}-\ref{eq:20b}). We only notice that in the limit
$\gamma\gg \beta$ our expression (\ref{eq:26}) becomes
\beq
{\cal P}(\nu_e\rightarrow \nu_e) \simeq 1 \label{eq:32}
\eeq
which means   no oscillations in this limit(see also eq(91) of
ref.{\cite{12}}).

Returning back to our new method we repeat that in the case of small mixing
angles for a resonance to occur, eqs(\ref{eq:12},\ref{eq:18}) impose a
stringent condition on the properties of the neutrinos.  Since in the solar
neutrino experiments the sun-earth distance $x$ is fixed we expect that the
resonance will occur for special values of $x_0$.  To test this we will
consider a reasonable model for the solar electron density employed
by Lim and Marciano $^{\cite{18}}$, i.e. we take
\beq
\frac{\rho_e(x)}{\rho_0} =
\left\{ \begin {array}{lc}
 1-a\frac{|\vec{x}|}{R_{\odot}},\,&\, 0<\mid{\vec{x}}\mid <
k_{1}R_{\odot}\\
 (1-a k_1)\, exp \{c(k_1 -\frac{\mid\vec{x}\mid}{R_{\odot}})\}\,&\,
\mid{\vec{x}}\mid >k_1 R_{\odot} \label{eq:33}
\end{array} \right\}
\eeq
%
with
$k_1=0.2$,    $a  =10/3$,  and $c =100/9$.
The value of $\rho_0$, absorbed in the definition of $t_0$
(see \ref{eq:15}) was chosen
\beq
\rho_0 = 6\times 10^{25} cm^{-3}\label{eq:34}
\eeq
With the above parameters we have perfomed calculations of
${\cal P}(\nu_e\rightarrow \nu_e)$ as given by eq.(\ref{eq:9}) near the
resonance  (\ref{eq:12}) for various values of  $x_0$.  The obtained results
for $sin^22\vartheta=2 m_e/m_{\mu} \approx 1.1\times 10^{-2}$ and  $x_0=0$,
$x_0=0.1 R_{\odot}$ and $x_0=0.2 R_{\odot}$  are shown in
figs (1-3).  The corresponding values of $E_{\nu}/\delta m^2$ in units of
$MeV-eV^{-2}$ are also indicated.
It is clear that the oscillation probability becomes negligible for regions
away from the resonance.  Our results are similar to those obtained for
$x_0=0$ in the context of the earlier treatment by Rosen and Gelb except
that:

{\it i)} The value of $E_{\nu}/\delta {m^2}$ at resonance is in our
approach  about a factor of 2 smaller.  This, however, can be accounted
for by the fact that they use a different electron density function.

{\it ii)} The width of our plots of oscillation probability as a function
of  $E_{\nu}/\delta m^2$ is quite a bit narrower.

We are currently analysing the solar neutrino data employing our new
formalism for a number of solar density profiles.  Detailed results will
be published elsewhere.  At present we note that for suitable values of
$E_{\nu}/\delta m^2$ the resonance condition for a small mixing angle is
satisfied in a small region in the $x_0$ - space.  One must appropriately
integrate over $x_0$ $^{\cite{15}}$.  If one does this, one expects
that even though some neutrinos may arrive at earth with maximal
 oscillation probability as given by eqs (\ref{eq:13a}) and
 (\ref{eq:13b}) the fraction of such
 neutrinos will be small.  A rough estimate can be given as follows
\begin{eqnarray}
\frac{N_{\nu}(res.)}{N_{\nu}(tot.)}&=&
\frac{\int {\cal P}(\nu_e\rightarrow \nu_e)
\rho(x_0)dx_0}{\int \rho(x_0)dx_0}
\end{eqnarray}
which in the interesting case of small mixing, i.e.
${x} sin 2\vartheta \ll {\ell}$ one finds
\beq
\frac{N_{\nu}(res.)}{N_{\nu}(tot.)}
\simeq  1 - \frac{x^2}{\ell^2} sin^2 2\vartheta \label{eq:35}
\eeq
If this is borne out by more detailed calculations in which integration
over the neutrino energy is also preformed $^{\cite{16,17}}$  it  will
imply  that in fact the medium may have a negligible effect on neutrino
oscillations.

In conclusion we have presented a novel way of treating neutrino
oscillations in a medium which is both simpler and more accurate than the
traditional perturbative approach.
 Our method can be easily extended in the case of
the three generations.  Our results given above are not expected to be
drastically altered by such an extension.

\vspace*{3mm}
{\bf  Acknowledgement :}
{\it This research was partially supported by the   EEC grant
SCI-0221-C(TT)}.


\newpage


\newpage

\vspace*{2cm}

{\bf Figure Captions}

\vspace*{1cm}

{\bf Figure 1}: Oscillation probability at Earth for a  neutrino created
at the center of the sun ($x_0=0$), a)
as a function of the neutrino energy $E_{\nu}$ devided by
$\delta m^2=m_{\nu_{\alpha}}^2-m_{\nu_{e}}^2$ and, b)
as a function of the distance
from the sun.

\vspace*{1cm}

{\bf Figure 2}: same as in figure 1 but for $x_0=0.1 R_{\odot}$

\vspace*{1cm}

{\bf Figure 3}: same as in figure 1 but for $x_0=0.2 R_{\odot}$

\end{document}